**DNA topology dictates strength and flocculation in DNA-microtubule composites**

Karthik R. Peddireddy[1], Davide Michieletto[2,3], Gina Aguirre[1], Jonathan Garamella[1], Pawan Khanal[1], and Rae M. Robertson-Anderson[1,*]

[1]*Department of Physics and Biophysics, University of San Diego, 5998 Alcala Park, San Diego, CA 92110, United States*

[2]*School of Physics and Astronomy, University of Edinburgh, Peter Guthrie Tait Road, Edinburgh, EH9 3FD, UK,* [3]*MRC Human Genetics Unit, Institute of Genetics and Molecular Medicine University of Edinburgh, Edinburgh EH4 2XU, UK*

**Abstract**

Polymer composites are ubiquitous in biology and industry alike, owing to their emergent desirable mechanical properties not attainable in single-species systems. At the same time, polymer topology has been shown to play a key role in tuning the rheology of polymeric fluids. However, how topology impacts the rheology of composites remains poorly understood. Here, we create composites of rigid rods (microtubules) polymerized within entangled solutions of flexible linear and ring polymers (DNA). We couple linear and nonlinear optical tweezers microrheology with confocal microscopy and scaled particle theory to show that composites of linear DNA and microtubules exhibit a strongly non-monotonic dependence of elasticity and stiffness on microtubule concentration due to depletion-driven polymerization and flocculation of microtubules. In contrast, composites of ring DNA and microtubules show a much more modest monotonic increase in elastic strength with microtubule concentration, which we demonstrate arises from the increased ability of rings to mix with microtubules.

**Introduction**

Polymer composites, multi-component materials consisting of a continuous polymer matrix reinforced with fillers,[1-5] are of wide-spread importance in the natural and commercial world, with applications ranging from aerospace engineering to drug delivery[6-12]. Commercial interest stems from the fact that composites often exhibit emergent rheological and structural properties that are superior to those of the constituent materials[2,4,5,13-20]. For example, composites of stiff and flexible polymers can exhibit enhanced strength, and stiffness compared to single-component systems while concomitantly reducing weight[21-24]. In biological cells, a network of stiff and semiflexible protein filaments (i.e. cytoskeleton), forms in the presence of a dense solution of flexible and folded macromolecules such as nucleic acids and polysaccharides (i.e. cytoplasm). Mechanical interactions between the cytoplasm and cytoskeleton have been shown to be critical to the cell life cycle[25-28]. Further, previous in vitro studies have shown that cytoskeletal composites exhibit emergent properties such as stress stiffening and mechanomemory[13,29,30] due to entropically-driven polymer rearrangement.

Despite the wide-spread interest and applicability of composites, the role that polymer topology and end-closure play in composite rheology remains largely unexplored[31,32]. At the same time, polymer topology has been shown to play a primary role in dictating the rheological properties of entangled polymers[13,32-36]. For example, linear polymers more effectively form entanglements and undergo affine deformation compared to ring polymers,[33,34] resulting in significantly enhanced elasticity and shear thinning. However, ring polymers can become threaded by surrounding polymers which can drastically slow relaxation timescales and increase viscoelasticity[14,37-45]. Further, ring-linear polymer blends have been shown to exhibit increased elasticity, stress-stiffening, and relaxation timescales compared to their single-topology counterparts[14,32,46]. However, how polymer topology impacts the rheology and structure of polymer composites remains an open question.

Here, we create composites of stiff microtubules and flexible DNA molecules of varying topologies. We polymerize varying concentrations of tubulin into microtubules (MT) in the presence of entangled linear (L) or ring (R) DNA and determine the roles that DNA topology and tubulin concentration play in the linear and nonlinear microrheological properties as well as the structure of DNA-MT composites (Fig 1). We show that DNA topology plays a prominent role in microtubule polymerization, network formation, and bundling, which ultimately dictates the mechanical response of the composites. Notably, microtubules increase the rigidity of entangled linear DNA ~10-fold at low MT concentrations but upon subsequent increases in microtubules the moduli steadily decrease. This surprising non-monotonic dependence, which we show arises from microtubule flocculation, is non-existent for ring DNA which shows a steady increase in rigidity with increasing MT concentration.

**Results and Discussion**

We first determine the dependence of DNA topology and microtubule concentration on linear viscoelasticity which we extract from thermal oscillations of embedded trapped microspheres (Methods, Figs 1, 2). As shown in Figure 2a, the elastic modulus $G'(\omega)$ of R-MT composites increases monotonically with increasing MT concentrations over the entire frequency range, indicating increased elasticity. Further, at low MT concentrations, $G'(\omega)$ exhibits power-law scaling of ~0.3, which is reduced to ~0.1 for [MT]$\geq$ 5 μm which is lower than the previously reported value of 0.17 for 10 μM MT solutions[47]. At these higher MT concentrations, $G'(\omega)$ is nearly frequency-independent and we can approximate an elastic plateau $G^0$ from the high-frequency values (Fig 2a). The increased magnitude and decreased frequency-dependence of $G'(\omega)$ are both signatures of increased elasticity and connectivity, as one may expect given the increased density of stiff polymers. However, the discrete shift from ~0.3 to 0.1 scaling, rather than a gradual decrease, suggests a potential phase transition, as we discuss below.

In stark contrast to R-MT composites, we observe a surprising non-monotonic dependence of *G′(ω)* on MT concentration in L-MT composites. As MT concentration increases from 0 to 2 μM *G′(ω)* increases by an order of magnitude followed by a subsequent decrease as MT concentration increases to 7.5 μM. L-MT composites also show reduced frequency-dependence of *G′(ω)* compared to linear DNA that is most apparent for the lowest MT concentration. This counterintuitive result suggests that DNA-MT composites are not simple entangled polymer composites in which increases in the stiff component increases the elastic response. Of note, despite the decrease in elastic response at higher MT concentrations for L-MT composites, at any given MT concentration, the elastic response of R-MT composites is significantly lower and the frequency-dependence is stronger than the linear counterparts, likely due to the reduced ability of rings to form entanglements[14,32]. This topology dependence can also be seen in Figure 2d in which the approximate $G^0$ values are plotted as a function of MT concentration. Using previously reported values of $G^0$ and scaling $G^0 \sim c^{1.4}$ for MT solutions we estimate $G^0 \approx 0.05$ Pa for a 2 μM MT solution[47]. In contrast, L-MT composites with 2 μM MT exhibits an order of magnitude higher $G^0$ value.

We also evaluate the complex viscosity, $\eta^*(\omega)$, as previous studies have shown that entangled ring and linear DNA both exhibit shear thinning $\eta^*(\omega) \sim \omega^{-\alpha}$, but ring DNA exhibits weaker thinning (smaller $\alpha$) compared to linear DNA due to their reduced ability to align with flow[14,32,33,35]. The addition of microtubules increases $\alpha$ to ~0.9 for both topologies, but while 2 μM MT is sufficient for the increase in L-MT composites, 5 μM MTs are required for R-MT composites. The delay in this increase for R-MT composites suggests that microtubule network formation may be more readily facilitated in the presence of linear DNA. At the same time, the apparent similarity in shear-thinning behavior for both composites suggests that microtubules may facilitate flow alignment of ring polymers.

To determine the robustness of the intriguing topology-dependent viscoelasticity of DNA-MT composites to large strains, we measure the nonlinear force response by displacing microspheres 30 μm ($\gamma = 6.7$) through the composites using strain rates up to 113 s$^{-1}$. As shown in Figure 3, nonlinear stress curves for all DNA-MT composites initially rise steeply, with an elastic-like strain dependence, before reaching a softer more viscous regime in which the slopes of the force curves are shallower. The addition of microtubules to both DNA topologies leads to an increase in the magnitude and slope of the force at large distances, suggesting that DNA-MT composites are more readily able to retain elastic memory in the nonlinear regime compared to DNA solutions which reach nearly completely viscous response at the end of the strain. However, the strong dependence of the force response on DNA topology and MT concentration seen in the linear regime is preserved. Namely, R-MT composites exhibit a weak monotonic increase in force as a function of MT concentration while L-MT composites exhibit a strong non-monotonic dependence.

To further quantify these trends and determine the strain-rate dependence, we compare the maximum force $F_{max}$ reached during strain for each DNA-MT composite as a function of strain rate. For reference, a fluid-like system should display a purely viscous response in which $F_{max} \sim \dot{\gamma}^1$, whereas a solid-like system should show minimal rate dependence ($F_{max} \sim \dot{\gamma}^0$). As shown, all composites exhibit power-law dependence $F_{max} \sim \dot{\gamma}^\beta$ for $\dot{\gamma} > 10$ s$^{-1}$, with exponents that depend on DNA topology and MT concentration. Pure DNA solutions exhibit a scaling $\beta \simeq 0.7$, independent of topology, in line with previous studies on ring-linear DNA blends[14] and tube extension models for flexible polymers in nonlinear regime[48-50]. While the addition of microtubules only modestly reduces the rate dependence for R-MT composites for all MT concentrations examined ($\beta \geq 0.6$), the addition of 2 μM MTs to linear DNA solutions reduces $\beta$ to ~0.4. However, upon subsequent increase in MT concentration, the scaling exponent increases to $\beta \geq 0.6$ for 7.5 μM MTs. This result suggests that in L-MT composites stiff microtubules can synergistically interact more readily with linear DNA compared to ring DNA to oppose flow-induced disentanglement. However, these synergistic interactions are most efficient at lower MT concentrations.

Following the strain, we measure the relaxation of the imposed stress and as the composite reorganizes to a new equilibrium configuration. As shown in Figure 4, while both pure DNA solutions relax all of their stress during the measurement window, all DNA-MT composites retain a non-zero force, indicative of elastic memory. Similar to previous studies on entangled ring and linear DNA, we fit each relaxation curve to a sum of three exponential decays, $F(t) = F_\infty + C_1 e^{-t/\tau_1} + C_2 e^{-t/\tau_2} + C_3 e^{-t/\tau_3}$, but here we include a non-zero terminal offset $F_\infty$. These fits, each with adjusted $R$-squared values of ≥0.99, yield three well-separated time constants ($\tau_1$, $\tau_2$, $\tau_3$; Fig 4c) and terminal offsets that are dependent on MT concentration and DNA topology (Figure 4a inset). All $\tau_i$ and $F_\infty$ values show minimal dependence on strain rate so the values we present in Figure 4 are averaged over $\dot{\gamma}$, with the error bars representing the rate dependence. The fractional amplitudes $\phi_i = C_i/(C_1 + C_2 + C_3)$ of each decay show much less dependence on MT concentration but do, however, display a rate dependence, which we present in Fig 4d,e.

$F_\infty$ values display a strong dependence on MT concentration, following the same trend as we see in Figs 2 and 3, with the largest sustained force occurring at the lowest MT concentration in L-MT composites, and with L-MT composites exhibiting higher values than their R-MT counterparts. The dependence of the time constants on MT concentration is substantially weaker than for $F_\infty$ but statistically significant in some instances. To understand the mechanisms underlying each time constant we first compare our measured constants for entangled linear DNA (no microtubules) to the principle relaxation timescales predicted by the reptation model for entangled linear polymers[51]: the entanglement time $\tau_e$ over which diffusing chain segments reach the edge of the tube, the disengagement time $\tau_D$ over which the polymer reptates out of its initial deformed tube, and the Rouse time $\tau_R$ over which elastic relaxation of the deformed polymer occurs. Within this framework, the predicted timescales for our linear DNA solution are $\tau_e \cong 0.1$ s, $\tau_R \cong 0.5$ s, and $\tau_D \cong 9$ s[51]. As shown (Fig 4c), $\tau_1$, $\tau_2$ and $\tau_3$ for linear DNA are comparable but slightly smaller than $\tau_e$, $\tau_R$, and $\tau_D$, respectively, likely due to nonlinear straining[52]. The fractional amplitudes (Fig 4d,e) further support this interpretation as we see a large drop in $\phi_3$ as $\dot{\gamma}$ increases while the amplitudes of the two fast timescales, corresponding to $\tau_e$ and $\tau_R$ increase. Faster rates more easily disrupt entanglements and thus reduce the propensity for relaxation via reptation, thereby increasing the relative contributions from $\tau_e$ and $\tau_R$.

Comparing ring and linear composites, L-MT composites show a nonmonotonic dependence of relaxation dynamics on MT concentration, with all time constants increasing >2x upon addition of 2 μM MTs, followed by subsequent reduction, while $\tau$ values for R-MT composites lack any significant dependence on microtubule concentration. This result suggests that stress relaxation in R-MT composites happens primarily through ring DNA rearrangements, with minimal interactions with the microtubules, possibly indicating reduced ability of rings to form entanglements with the microtubules. The topology-dependent fractional amplitudes (Fig 4d,e), further support this interpretation as at low strain rates, $\phi_3$ for R-MT composites is lower than for L-MT composites. A steeper drop in $\phi_3$ in L-MT further suggests that L-MT composites are richer in entanglement interactions as forced disentanglement is rate dependent. Further, while strains rates up to 113 s$^{-1}$ are possible in R-MT composites, the trapping strength could not withstand rates greater than 30 s$^{-1}$ in L-MT composites, indicating stronger entanglements and DNA-MT interactions.

To make sense of the topology dependence in DNA-MT composites and the surprising non-monotonicity in L-MT composites, we examine confocal micrographs of composites with rhodamine-labeled microtubules. Figure 5 shows representative images for all DNA-MT composites along with MT networks polymerized in the absence of DNA. Without DNA, tubulin polymerizes into disconnected branched clusters of microtubules that are heterogeneously distributed throughout the sample. As MT concentration increases these branched clusters grow and become more interconnected.

Typically, the addition of crowding agents or depletants enhances polymerization reactions due to entropically-driven depletion effects[53-55]. Namely, the surrounding crowders aim to maximize their entropy by driving the polymerizing monomers together to reduce the excluded volume that surrounds each monomer. Indeed, microtubule polymerization has been shown to be enhanced by macromolecular crowders such as polyethylene glycol and bovine serum albumin[56-58].

Surprisingly, we see an opposite effect in R-MT composites. The presence of ring DNA hinders microtubule polymerization. Namely, we observe no visible microtubules at 2 μM tubulin and very limited polymerization at 3.5 μM. Only at 5 μM and above does a percolated microtubule network emerge. This result is consistent with our linear microrheology results (Fig 2) that show a discrete shift in $G'(\omega)$ scaling at 5 μM MT. We further note that the networks that form at 5 μM and above are much more homogeneously distributed, with fewer clusters and branches compared to MTs alone, indicating that the entropic gain from mixing of the two species outweighs the entropic gain from depletion interactions. Previous studies have shown that end-closure of polymers can significantly increase the miscibility of polymer blends[59,60]. Namely, ring polymers can more readily mix with other polymeric species than their linear counterparts. This increased miscibility appears to transfer to composites with disparate species.

Opposite to our R-MT results, L-MT composites shows a percolated MT network at 2 μM tubulin that is significantly more connected and pervasive than for MTs alone. The enhanced network formation in the presence of linear DNA, likely driven by depletion interactions[61-63], explains the large increase in the force response and moduli for L-MT composites at 2 μM tubulin. Specifically, a percolated network of stiff microtubules provides a scaffold to reinforce the entangled DNA, while, at the same time, entanglements with the linear DNA provide elastic support to the MT network. The confocal images further shed light on the subsequent drop in force response we measure at higher MT concentrations. As the MT concentration increases more bundling of microtubules occurs, evidenced by brighter clusters with larger voids, which lowers the connectivity of the microtubule network. This reduced connectivity weakens the microtubule scaffold and reduces the interactions with the DNA, thereby providing less structural support.

To further characterize DNA-MT composite structure, we compute the spatial image autocorrelation $g(r)$ (Fig 5b). All $g(r)$ curves exhibit exponential decay rather than power-law behavior suggestive of micro-phase separation instead of a self-similar fractal structure[42]. By fitting each curve to $g(r) \propto e^{r/\xi}$ we can extract a characteristic correlation length $\xi$ that describes the structure of the network. As shown, both DNA topologies decrease the correlation length of the MT network compared to when tubulin is polymerized without DNA. R-MT composites have the smallest ξ values for all concentrations >2 μM (in which no network forms). At any given concentration, the size of the microtubules are smaller than in the other cases, and, when a connected network forms, the mesh size is smaller than in the other cases as it is comprised of individual homogenously distributed filaments rather than bundles or clusters. ξ values for L-MT composites are significantly higher than for R-MT composites and increases from 2 μM to 5 μM. As fully connected networks are evident at all concentrations, this increase in ξ represents an increase in mesh size as the microtubules flocculate together. This flocculation in turn weakens the network by reducing the entanglements and connections with the DNA – thus explaining the drop in force we measure as tubulin concentration increases above 2 μM.

To explain these intriguing observations we use Scaled Particle Theory[64] to compute the phase diagram of a solution of rod-like colloids (MTs) and flexible coils (DNA) (Fig 6). Within this framework, thoroughly described in SI, the DNA works as a depletant and induces an isotropic-to-nematic transition for the MTs for certain critical values of DNA and MT volume fractions, as shown by the binodals in Fig 6. The region of phase space between the binodals represents the coexistence of isotropic and nematic phases of the MTs, i.e. flocculation. Specifically, small groups of MTs nematically align with one another but these bundles or 'flocs' are still uniformly and isotropically distributed, as shown in Fig 5.

Intriguingly, the presence of the DNA depletant significantly widens the coexistence region from that of a simple (Onsager) solution of rod-like colloids[65,66], demonstrating an emergent feature of flexible-stiff polymer composites. Further, in agreement with our experiments, our calculations show that ring DNA depletants are substantially less effective at inducing flocculation of MTs, as evidenced by the smaller coexistence regime and the higher MT concentrations required to induce the phase transition. This smaller coexistence regime further shows that the nematic phase of MTs is less dense in the presence of ring DNA compared to linear DNA. This effect implies that the modulation of density in an R-MT composite is weaker than that for L-MT composites, and is therefore less likely to impact the mechanical stability and strength of the MT network, as evidence in our experiments by a weaker dependence of the force response on MT concentration.

**Conclusion**

In conclusion, our optical tweezers microrheology and confocal microscopy studies on DNA-microtubule composites, combined with our scaled particle theory calculations, show that subtle changes in polymer topology (free or closed ends) can have a dramatic effect on the structure and mechanics of polymer composites. Specifically, linear DNA promotes microtubule network formation and flocculation while ring DNA hinders it, which leads to a substantially larger force response in L-MT composites compared to R-MT composites, as well as a unique non-monotonic dependence of elastic strength on tubulin concentration. Our results shed important new light on the role that topology plays in the rheology and structure of polymer composites, which has broad reaching implications in biology, chemical engineering and materials applications. The realization of polymer composites through in-situ polymerization of fillers has further advantages in wide variety of applications such as in vivo tissue engineering and in-situ polymerization for 3D printing.

**Methods**

Many of the materials and methods are described in the preceding sections and in the captions of Figs 1-6. More detailed descriptions of all experimental materials and methods as well as theoretical calculations are included in the Supplementary Information.

**Acknowledgements**

The authors acknowledge financial support from Air Force Office of Scientific Research (AFOSR-FA9550-17-1-0249). The authors also acknowledge Prof. Jennifer L Ross (Syracuse University) for useful discussions regarding microtubule preparation and handling.

**Author contributions**

R.M.R.-A. conceived the project, guided the experiments, interpreted the data, and wrote the manuscript. K.R.P. designed and performed the experiments, analyzed and interpreted the data and wrote the manuscript. D.M. performed theoretical calculations, analyzed and interpreted results, and helped write the manuscript. G.A, J.G. and P.K prepared and characterized DNA solutions used in experiments.

**Competing interests**

The authors declare no competing interests.

**Materials and Correspondence**

Correspondence and requests for materials should be addressed to R.M.R-A. (*Email: randerson@sandiego.edu)

**ORCID**

Karthik Reddy Peddireddy: 0000-0003-4282-5484
Davide Michieletto: 0000-0003-2186-6869
Jonathan Garamella: 0000-0002-5704-7606
Pawan Khanal: 0000-0003-4126-7951
Rae M. Robertson-Anderson: 0000-0003-4475-4667

**Figures**

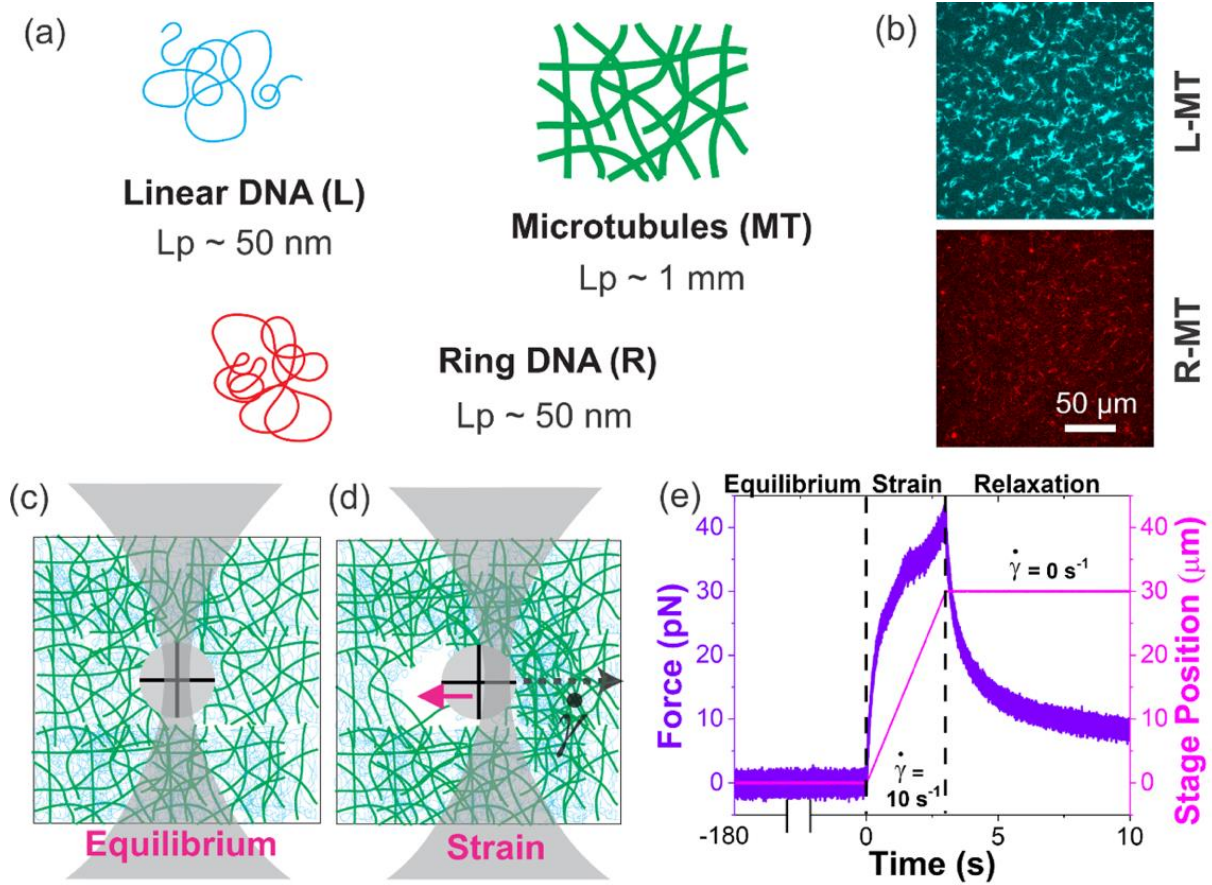

**Figure 1. Elucidating the microrheological properties of composites of rigid microtubules and flexible ring and linear DNA.** (a) Cartoons of linear (L, blue) and ring (R, red) DNA of identical contour lengths (115 kbp, 39 µm) and microtubules (MT, green). (b) Confocal micrographs of composites of 5 µM rhodamine-labeled microtubules polymerized in 0.65 mg/ml solutions of linear (top, L-MT) and ring (bottom, R-MT) DNA (unlabeled). The scale bar applies to both micrographs. (c-e) Cartoons of linear (c) and nonlinear (d) optical tweezers microrheology. A microsphere of radius $R$=2.25 µm (grey circle) embedded in an L-MT composite is trapped using a focused laser beam (grey). (e) Thermal oscillations of the bead in equilibrium are used to determine the linear viscoelastic moduli via the generalized Stokes-Einstein relation (see Methods). To measure the nonlinear rheological response, the same optically trapped bead is displaced 30 µm ($\gamma$=6.7) through DNA-MT composites at speeds $v = 2.5–120$ µm/s ($\dot{\gamma} = 3v/\sqrt{2}R = 2.4 – 113$ s$^{-1}$). Following strain, the composite is allowed to relax, and the microsphere returns to the trap center. Stage position (magenta) and force exerted on the trapped bead (violet) during Equilibrium (180 s), Strain (0.25-12 s), and Relaxation (8-20 s) (delineated by dashed lines) are recorded at 20 kHz. Data shown is for a $v = 10$ µm/s strain exerted on an L-MT composite with 2 µM microtubules.

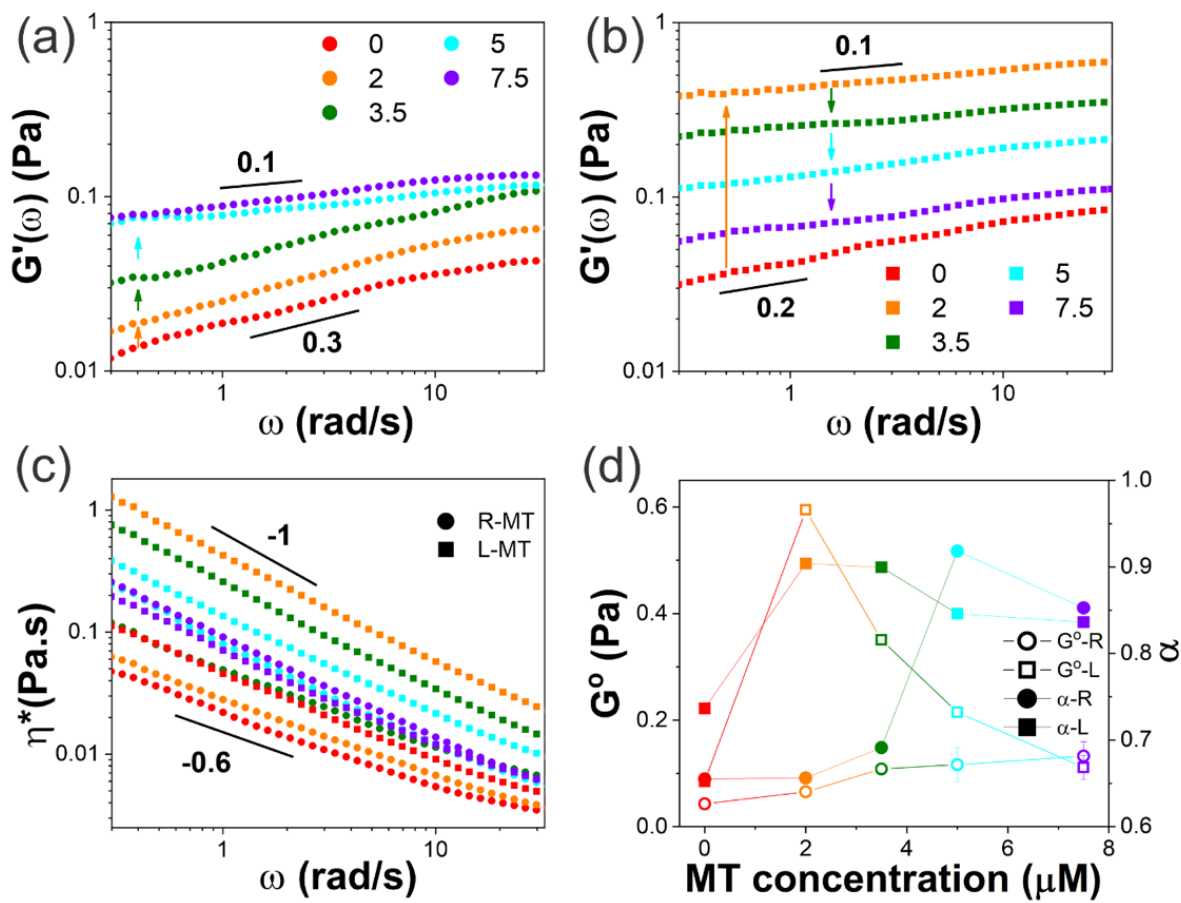

**Figure 2. Linear frequency-dependent viscoelastic moduli of DNA-MT composites exhibit strong dependence on DNA topology.** (a-b) Frequency-dependent elastic moduli $G'(\omega)$ for DNA-MT composites with ring (a, circles) and linear (b, squares) DNA and varying microtubule concentrations (shown in μM of tubulin in legend). Arrows point in the direction of increasing MT concentration to guide the eye. All $G'(\omega)$ curves approach elastic plateaus at high frequencies with a strong non-monotonic dependence on MT concentration for linear DNA. (c) Complex viscosity $\eta^*(\omega)$ for R-MT (circles) and L-MT (squares) composites show shear-thinning $\eta^*(\omega) \sim \omega^{-\alpha}$ with exponents that depend on DNA topology. Black lines denote power-laws with exponents listed. (d) Elastic plateau modulus $G^0$ (open symbols) determined from high-frequency $G'(\omega)$ values and shear-thinning exponent $\alpha$ (closed symbols) determined from power-law fits to $\eta^*(\omega)$, plotted for R-MT (circles) and L-MT (squares) composites as a function of MT concentration.

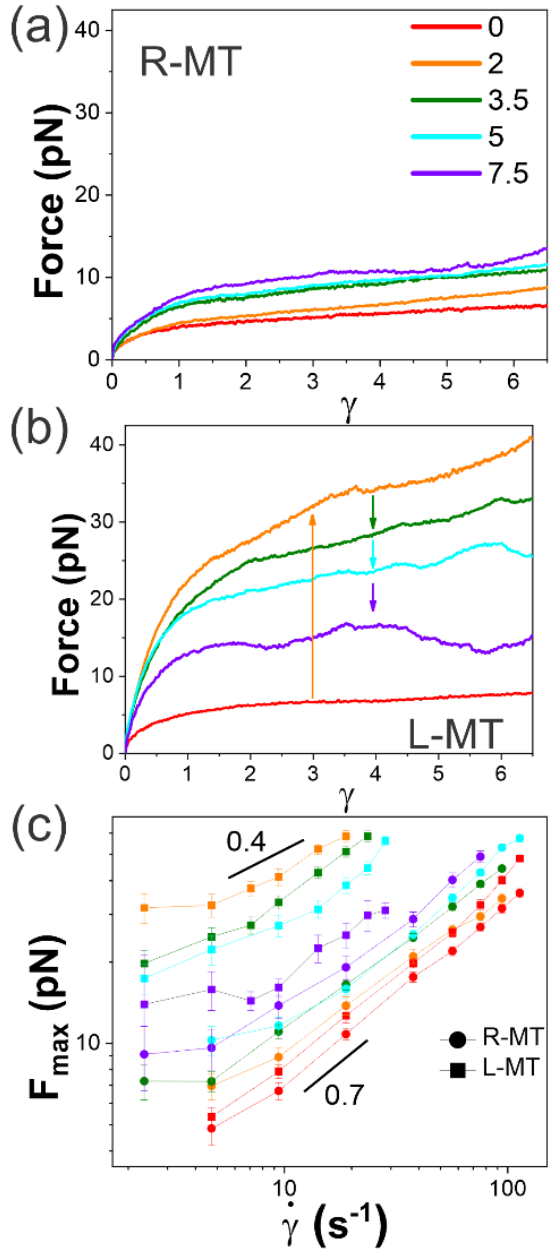

**Figure 3. The nonlinear force response of DNA-MT composites exhibits a complex dependence on DNA topology, microtubule concentration, and strain rate**. (a-b) Measured force in response to strain of rate $\dot{\gamma}=9.4$ s$^{-1}$ for composites with ring (a) or linear (b) DNA and varying MT concentrations shown in μM of tubulin in the legend. Arrows point in the direction of increasing MT concentration to guide the eye. Similar to the linear regime, L-MT composites exhibit greater force response than R-MT composites and a strong non-monotonic dependence on microtubule concentration that is lacking in R-MT composites. (c) Maximum force reached during strain $F_{max}$ versus strain rate $\dot{\gamma}$ for R-MT (circles) and L-MT (squares) composites with varying MT concentrations shown in the legend in (a). Black lines represent power-law scaling with exponents shown. In general, composites with higher force responses have weaker dependence on strain rate, signifying a more elastic response. Error bars represent the standard error from 15 trials.

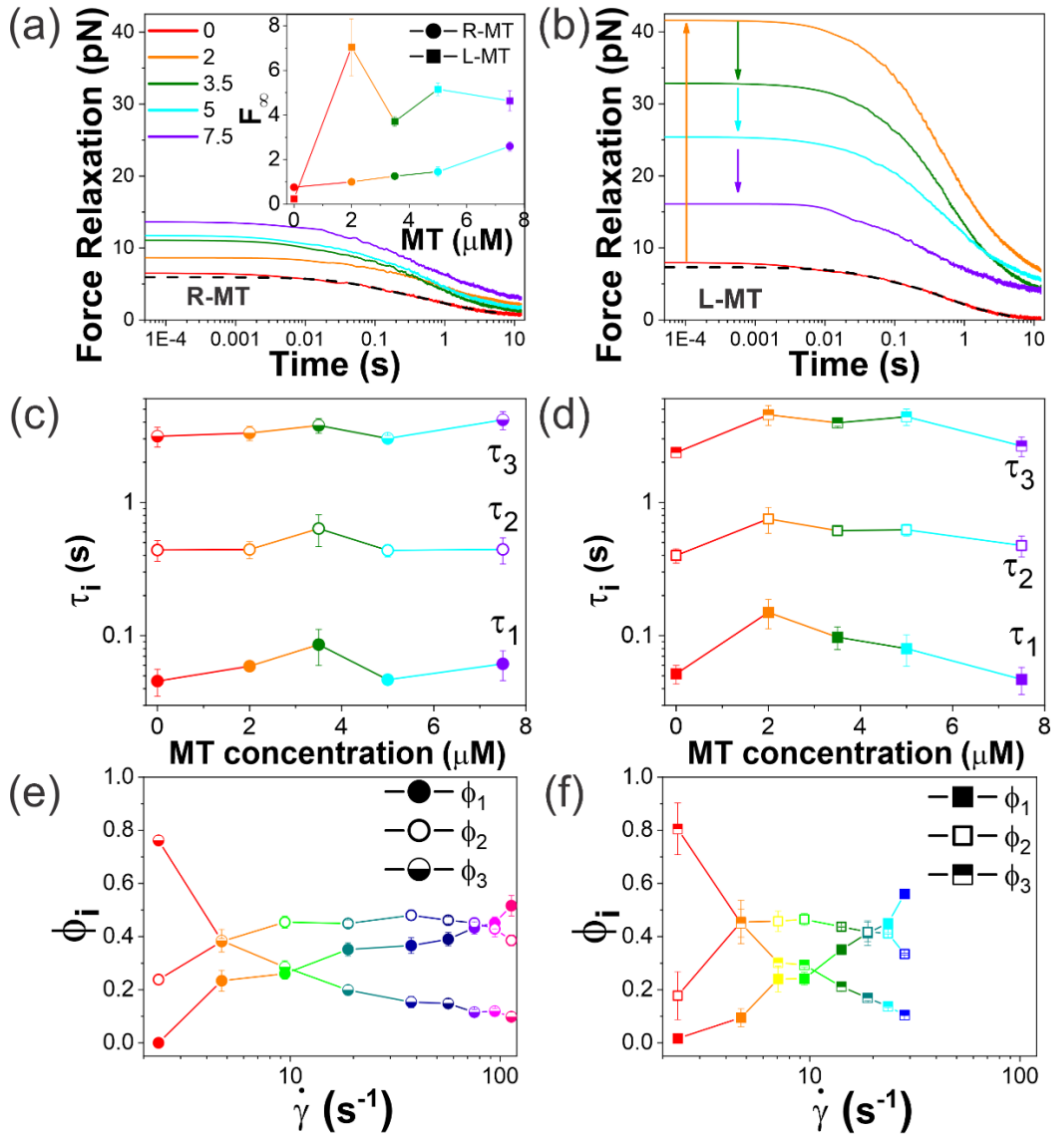

**Figure 4. DNA-MT composites exhibit multi-mode relaxation and sustained memory following nonlinear straining**. (a-b) Force relaxation of composites with ring (R-MT, a) or linear (L-MT, b) DNA and varying MT concentrations (shown in μM in legend) following a $\dot{\gamma}$=9.4 s$^{-1}$ strain. Each relaxation curve is fit to a sum of three exponential decays: $F(t) = F_\infty + C_1 e^{-t/\tau_1} + C_2 e^{-t/\tau_2} + C_3 e^{-t/\tau_3}$. Sample fits are shown as black dashed lines. (Inset) DNA-MT composites exhibit sustained elasticity as shown by the non-zero force maintained at the end of the relaxation phase $F_\infty$, shown averaged over all $\dot{\gamma}$ as a function of MT concentration. (c-d) Time constants $\tau_1$, $\tau_2$, and $\tau_3$, determined from fits and averaged over all $\dot{\gamma}$, as a function of MT concentration for R-MT (circles, c) and L-MT (squares, d) composites. (e-f) Corresponding fractional amplitudes $\phi_1$ [$= C_1/(C_1 + C_2 + C_3)$] $\phi_2$, and $\phi_3$ determined from fits, averaged over all MT concentrations and plotted versus $\dot{\gamma}$ for R-MT (circles, e) and L-MT (squares, f) composites. Fractional amplitudes for both R-MT (circles) and L-MT (squares) composites show that fast relaxation modes ($\tau_1$ and $\tau_2$) become increasingly dominant at high strain rates whereas the slowest mode ($\tau_3$) dominates at low strain rates. Fractional amplitudes for $\dot{\gamma}$>30 s$^{-1}$ are not available for L-MT composites as the composite resistive force exceeds the trapping force.

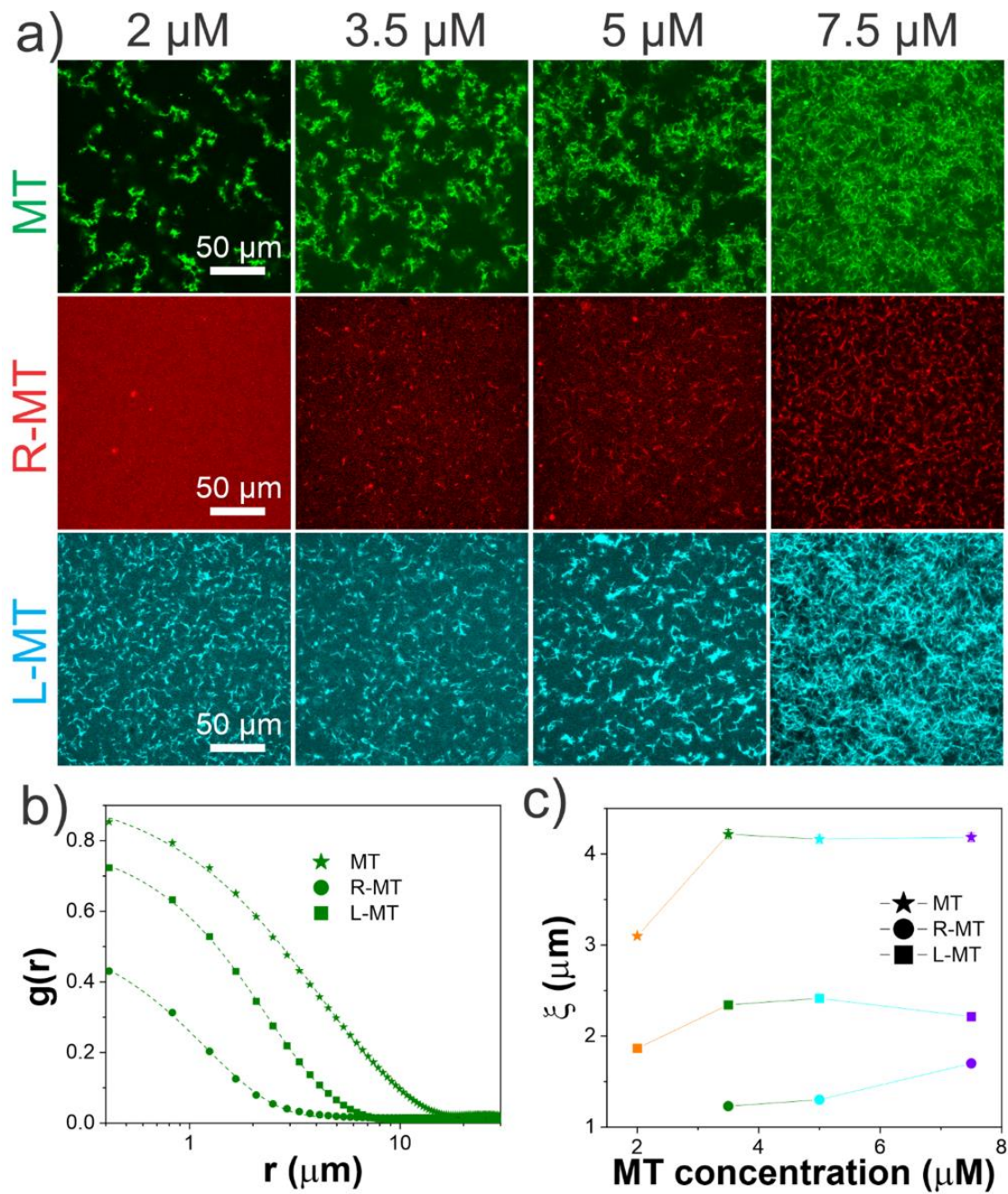

**Figure 5. DNA topology dictates the degree of polymerization and flocculation of microtubules in DNA-MT composites.** (a) Confocal micrographs of microtubules polymerized from rhodamine-labeled tubulin dimers of varying concentrations (listed above each row in μM) in buffer (MT, top, green), ring DNA solutions (R-MT, middle, red), and linear DNA solutions (L-MT, bottom, cyan). (b) Average spatial image autocorrelation curves $g(r)$ computed from confocal images for [MT]=3.5 μM (see Fig S2 for complete set of autocorrelation curves). (c) Structural correlation lengths ξ determined from fits of the autocorrelation curves to $g(r) \propto e^{r/\xi}$ for all cases shown in (a).

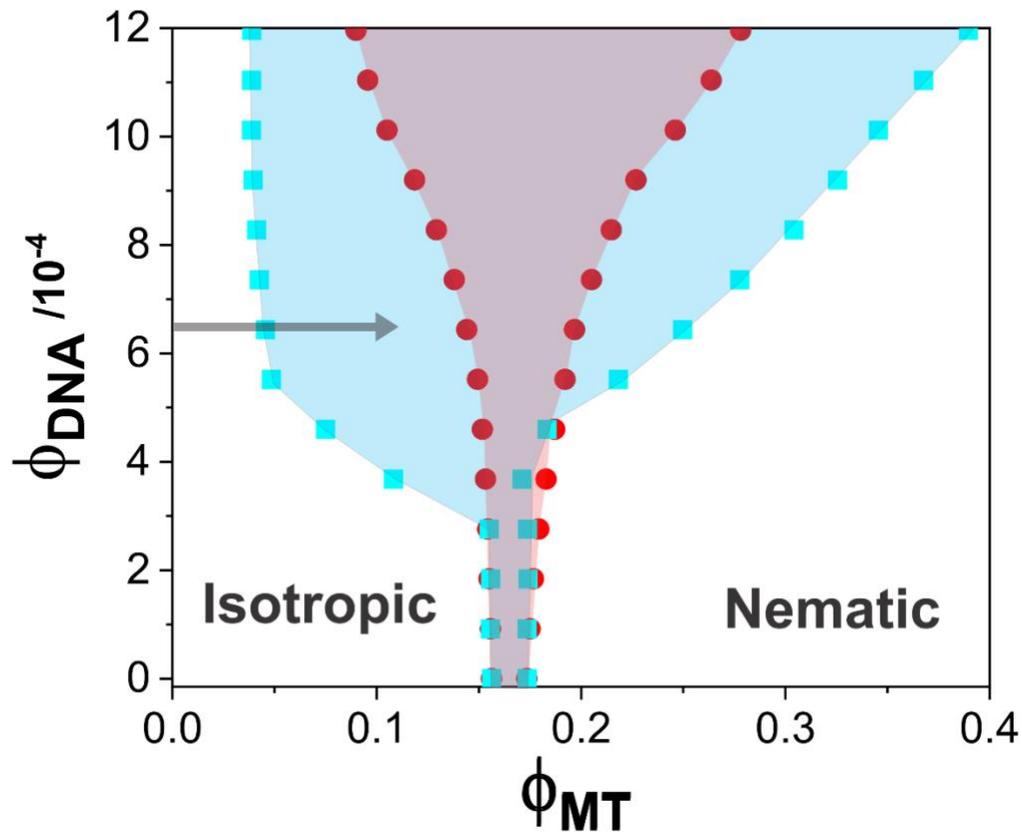

**Figure 6. Depletion-induced isotropic-to-nematic transition explains the non-monotonic mechanical response and flocculation of microtubules in DNA-MT composites.** Phase diagram showing results using scaled particle theory[64] for a composite of rod-like particles (MTs) and flexible polymer coils (DNA) for varying volume fractions $\phi_{MT}$ and $\phi_{DNA}$ (see SI for thorough description of theory and calculations). The points mark the binodal lines separating the isotropic and nematic phases for the microtubules in the presence of ring (red) or linear (cyan) DNA. The shaded coexistence regions between the binodals indicate the region of the phase space in which MT flocculation occurs for ring (red) and linear (cyan) DNA. Note that the coexistence region is significantly smaller for ring DNA and shifted towards higher MT concentrations, as seen in experiments (Fig 5).

# DNA topology dictates strength and flocculation in DNA-microtubule composites

Karthik R. Peddireddy[1], Davide Michieletto[2,3], Gina Aguirre[1], Jonathan Garamella[1], Pawan Khanal[1], and Rae M. Robertson-Anderson[1,*]

[1]Department of Physics and Biophysics, University of San Diego, 5998 Alcala Park, San Diego, CA 92110, United States

[2]School of Physics and Astronomy, University of Edinburgh, Peter Guthrie Tait Road, Edinburgh, EH9 3FD, UK, [3]MRC Human Genetics Unit, Institute of Genetics and Molecular Medicine University of Edinburgh, Edinburgh EH4 2XU, UK

**SUPPLEMENTARY INFORMATION**

**Section 1. Experimental Section**

**Figure S1. Linear frequency-dependent viscoelastic moduli for 0.65 mg/ml solutions of 115 kbp linear (squares) and ring (circles) DNA**

**Figure S2. Average spatial image autocorrelation curves of DNA-MT composites**

**Section 2. Theory for phase behavior of a composite of stiff and flexible polymers**

**Section 1. Experimental Section**

*DNA:* Double-stranded 115 kilobasepair DNA is prepared by replication of cloned bacterial artificial chromosomes (BACs) in *E. Coli*, followed by extraction, purification, and concentration using custom-designed protocols described elsewhere[1,2]. Supercoiled BAC DNA constructs are enzymatically treated with MluI or topoisomerase I (New England Biolabs) to convert the supercoiled constructs to linear (L) or ring (R) topology, respectively. Both linear and ring DNA stock solutions are suspended in TE10 buffer (10 mM Tris-HCl (pH 8), 1 mM EDTA, 10 mM NaCl) at concentrations ($c$) of 1.9 mg/ml and 1.1 mg/ml, respectively. DNA concentration for experiments is fixed at $c=0.65$ mg/ml, which corresponds to $2.5c_e$ where $c_e$ is the critical entanglement concentration[3]. This concentration translates to $N_e \approx 5$ entanglements for linear DNA, computed using traditional Doi-Edwards theory, which predicts $N_e = (4/5)cRT/(MG^0)$, where $M$ is the molecular weight of the polymer and $G^0$ is the plateau modulus[4]. $G^0$ is obtained from linear microrheology experiments (Figure S1).

*Microtubules (MT):* Porcine brain dark tubulin (T240) and rhodamine-labeled tubulin (TL590M) are obtained from Cytoskeleton. Tubulin stock solutions (45.5 μM) containing 9:1 dark:labeled tubulin in PEM100 buffer (100 mM PIPES (pH 6.8), 2 mM $MgCl_2$, 2mM EGTA) are flash-frozen in liquid nitrogen and stored at -80°C. To form DNA-microtubule composites, microtubules are grown directly in DNA solutions by adding 2 mM GTP, to enable polymerization, and 10 μM Taxol to stabilize polymerized microtubules.

*Sample preparation:* DNA and tubulin are mixed slowly and thoroughly using wide-bore pipette tips, to prevent shearing of DNA, then introduced into sample chambers through capillary action. Sample chambers (2 cm x 0.3 cm x 0.01 cm) are made with a microscope glass slide and coverslip separated by two layers of double-sided tape and hermetically sealed with epoxy. To polymerize tubulin dimers and form DNA-microtubule samples, sample chambers are incubated at 37°C for two hours, resulting in repeatable and reliable polymerization of microtubules in the DNA solutions. As an alternative method to forming composites, we also polymerized microtubules prior to adding to DNA solutions and loading into the sample chamber. However, this method resulted in random flow alignment of microtubules that occurred when introducing the DNA-MT mixtures into the sample chambers, so the data is not shown here.

*Microrheology:* For microrheology experiments, a trace amount of polystyrene beads (Polysciences, Inc.) of radius $R$=2.25 μm are added to the DNA-tubulin mixtures prior to microtubule polymerization. Beads are coated with Alexa-488 BSA to prevent DNA adsorption and for fluorescence visualization. 0.1 wt% Tween 20 is added to reduce DNA and bead adsorption to the sample chamber walls. An oxygen scavenging system (45 µg/mL glucose, 43 µg/mL glucose oxidase, 7 µg/mL catalase, and 5 µg/mL β-mercaptoethanol) is added to inhibit photobleaching.

We use optical tweezers microrheology to determine linear and nonlinear rheological properties of the DNA-MT composites (Figure 1). Details of the experimental procedures and data analysis, briefly summarized below, have been described in detail in refs [5,6]. The optical trap, built around an Olympus IX71 epifluorescence microscope, is formed from a 1064 nm Nd:YAG fiber laser (Manlight) focused with a 60x 1.4 NA objective (Olympus). Forces exerted by the DNA-MT composites on the trapped beads are determined by recording the laser beam deflections via a position sensing detector (Pacific Silicon Sensors) at 20 kHz. The trap is calibrated for force measurement using the stokes drag method.

Linear viscoelastic properties are determined from thermal fluctuations of a trapped microsphere, measured by recording the associated laser deflections for 180 seconds at 20 kHz. Linear viscoelastic moduli, i.e. the elastic modulus $G'(\omega)$ and the viscous modulus $G''(\omega)$, were extracted from the thermal fluctuations using the generalized Stokes-Einstein relation (GSER) as described in ref [7]. The procedure requires extracting the normalized mean-squared displacements ($\pi(\tau) = <r^2(\tau)>/2<r^2>$) of the thermal forces, averaged over all trials, which is then converted into the Fourier domain via:

$$-\omega^2 \pi(\omega) = \left(1 - e^{-i\omega\tau_1}\right)\frac{\pi(\tau_1)}{\tau_1} + \dot{\pi}_\infty e^{-i\omega t_N} + \sum_{k=2}^{N}\left(\frac{\pi_k - \pi_{k-1}}{\tau_k - \tau_{k-1}}\right)\left(e^{-i\omega\tau_{k-1}} - e^{-i\omega\tau_k}\right),$$

where $\tau$, 1 and N represent the lag time and the first and last point of the oversampled $\pi(\tau)$. $\dot{\pi}_\infty$ is the extrapolated slope of $\pi(\tau)$ at infinity. Oversampling is done using the MATLAB function PCHIP. $\pi(\omega)$ is related to viscoelastic moduli via:

$$G^*(\omega) = G'(\omega) + iG''(\omega) = \left(\frac{k}{6\pi R}\right)\left(\frac{1}{i\omega\pi(\omega)} - 1\right),$$

where $R$ and $k$ represent the radius of the microsphere and trap stiffness. We computed the complex viscosity $\eta^*(\omega)$ via $\eta^*(\omega) = [(G'(\omega))^2 + (G''(\omega))^2]^{1/2}/\omega$.

Nonlinear microrheology measurements are performed by displacing a trapped microsphere, $x = 30\ \mu m$, through the sample at speeds of $v = 2.5 - 120$ μm/s using a piezoelectric nanopositioning stage (Mad City Laboratories) to move the sample relative to the microsphere. We convert the distance to strain via $\gamma = x/2R$, and convert speeds to strain rates via $\dot{\gamma} = 3v/\sqrt{2}R$ (2.4-113 s$^{-1}$) [8]. The strain of 6.7 is much higher than the critical value of 1 for nonlinearity and our chosen strain rates are higher than the terminal relaxation frequencies $\omega_T = \lim_{\omega \to 0} \omega G''/G'$ of all DNA-MT composites under investigation. For both linear and nonlinear measurements, all data is recorded at 20 kHz, and at least 15 trials are conducted, each with a new microsphere in an unperturbed location. Presented data is an average of all trials.

*Confocal microscopy:* Microtubules in DNA-MT composite networks are imaged using a Nikon A1R laser scanning confocal microscope with a 60× 1.4 NA objective. 10% of tubulin dimers comprising microtubules are rhodamine-labeled to enable imaging using a 561 nm laser with 561 nm excitation and 595 nm emission filters. DNA in composites is unlabeled. 512×512 pixel images (212 μm ×212 μm) are taken at 20 different locations in the sample for each DNA-MT composite. Spatial image autocorrelation analysis is performed on each image using a custom-written python script[9,10]. Autocorrelation curves, $g(r)$, for each image are obtained by measuring the correlation in intensity $I(r)$ as a function of distance $r$.[9] Specifically, $g(r)$ is computed from $I(r)$ via:

$$g(r) = \frac{F^{-1}(|F(I(r))|^2)}{[I(r)]^2}$$

where $F$ and $F^{-1}$ represent fast Fourier and inverse Fourier transforms. The correlation length ξ is obtained by fitting autocorrelation curves to $g(r) = Ae^{r/\xi}$.

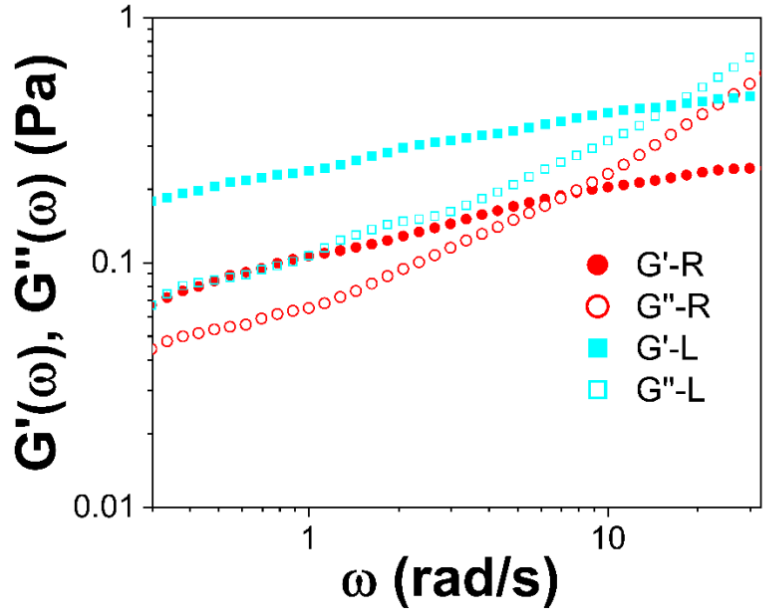

**Figure S1. Linear frequency-dependent viscoelastic moduli for 0.65 mg/ml solutions of 115 kbp linear (squares) and ring (circles) DNA.** Elastic (*G'(ω)*, closed symbols) and viscous moduli (*G''(ω),* open symbols) versus angular frequency *ω*.

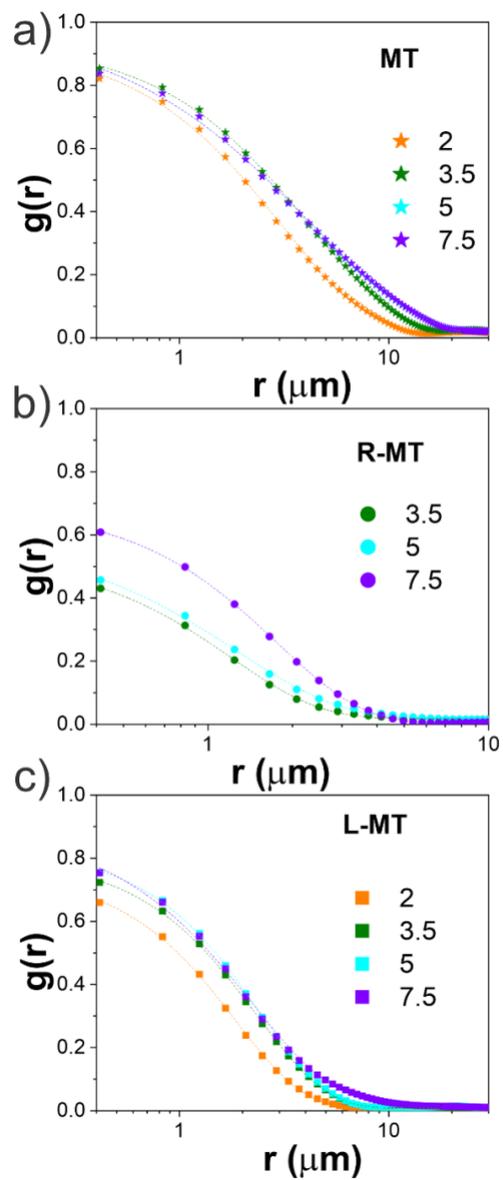

**Figure S2. Average spatial image autocorrelation curves of DNA-MT composites**. Each curve is an average of $g(r)$ curves computed for 20 different images of rhodamine-labeled microtubules taken at different locations of 2 independent samples for (a) microtubules without DNA, (b) R-MT composites and (c) L-MT composites with varying tubulin concentrations shown in $\mu$M in the legends.

**Section 2: Theory for phase behavior of a composite of stiff and flexible polymers**

The free energy of a solution of hard rod-like colloids and flexible polymers is

$$F = F_R(N_R, V, [f]) - \Pi(\mu_P)V_f$$

Where $F_R$ is the Helmoltz free energy of a system of $N_R$ rods in a volume $V$ and $\Pi = \mu_p n_p$ is the osmotic pressure exerted by a reservoir of polymer molecules, equal to the chemical potential multiplied by the number density of polymer molecules. Finally, $V_f$ is the free volume available to the polymer computed by the average volume available to polymers and $f$ the orientation distribution of the rods. This function captures the fact that the rods can be in an isotropic or nematic state and assumes a shape that minimizes the free energy.

*Free energy of rod colloids*

To express $F_R$ as a function of the density including corrections beyond the second virial coefficient we use the Scaled Particle Theory (SPT) description[11,12]:

$$\frac{F_R}{N_c k_B T} = \log n_c - \log(1 - \phi) + \sigma + X_2 \frac{\phi}{1 - \phi} + \frac{1}{2} X_3 \left(\frac{\phi}{1 - \phi}\right)^2 \qquad (1)$$

Where $n_c = N_c/V$ is the colloid particle density,

$$\phi = n_c \left(\frac{\pi}{6} D^3 + \frac{\pi}{4} D^2 L\right) \qquad (2)$$

their volume fraction. The coefficients $X_2$ and $X_3$ are given by

$$X_2 = 3 + \frac{3(\gamma - 1)^2}{3\gamma - 1} \rho[f]$$

$$X_3 = \frac{12\gamma(2\gamma - 1)}{(3\gamma - 1)^2} + \frac{12\gamma(\gamma - 1)^2}{(3\gamma - 1)^2} \rho[f]$$

Where $\gamma = (L + D)/D$ and the quantities

$$\sigma = \int d\Omega \, f \log(4\pi f)$$

$$\rho = \frac{4}{\pi} \int \int d\Omega \, d\Omega' \, ff' \sin(\theta)$$

And $\theta$ is the angle between two rod-like colloids. To simplify the calculations, we will use the ansatz that $f$ is a Gaussian distribution – this reduces the problem of finding the free energy minimizing orientation distribution of colloids to that of minimizing a variational parameter $\alpha$ defined via the function $f(\theta) = \frac{\alpha}{4\pi} e^{-\frac{1}{2}\alpha\theta^2}$ for $0 < \theta < \frac{\pi}{2}$ and $f(\theta) = \frac{\alpha}{4\pi} e^{-\frac{1}{2}\alpha(\pi - \theta)^2}$ for $\frac{\pi}{2} < \theta < \pi$. Importantly, using large $\alpha$ (nematic phase), one finds

$$\sigma = \sigma(\alpha) \simeq \log(\alpha) - 1 \tag{3}$$

and

$$\rho = \rho(\alpha) \simeq 4/\sqrt{\pi\alpha}. \tag{4}$$

In the isotropic phase, these terms are $\sigma = 0$ and $\rho = 1$.

*Free volume of polymers*

To obtain the free volume available to polymers, we employ Widom's particle insertion method, which relates the chemical potential of a polymer species to work needed to insert one such a polymer in the system. We thus write

$$\mu_p = k_B T \log \frac{N_p}{V_f} = k_B T \left( \log \frac{N_p}{V} - \log \frac{V_f}{V} \right)$$

Which also equals

$$\mu_p = k_B T \log \frac{N_p}{V} + W$$

Putting the two equations together one obtains

$$v = \frac{V_f}{V} = e^{-W/k_B T} = (1-\phi)\exp(-Ay - By^2 - Cy^3) \tag{5}$$

Where $y = \phi/(1-\phi)$ and

$$A = \frac{6\gamma}{3\gamma-1}q + \frac{3(\gamma+1)}{3\gamma-1}q^2 + \frac{2}{3\gamma-1}q^3$$

$$B = \frac{1}{2}\left(\frac{6\gamma}{3\gamma-1}\right)^2 q^2 + \left(\frac{6}{3\gamma-1} + \frac{6(\gamma-1)^2}{3\gamma-1}\rho[f]\right)q^3$$

$$C = \frac{2}{3\gamma-1}\left(\frac{12\gamma(2\gamma-1)}{(3\gamma-1)^2} + \frac{12\gamma(\gamma-1)^2}{(3\gamma-1)^2}\rho[f]\right)q^3$$

With $q = \sigma/D$ is the ratio between the diameter of the flexible polymer and the diameter of the rod-like colloid and $\gamma = (L+D)/D$[11].

Combining Eqs. (3-5) into (1) yields a semi-grand canonical free energy with one ($\alpha$) variational parameter that needs to be tuned to its free-energy-minimising value. To find it, we numerically compute the minimum of Eq. (1) as a function of $\alpha$ and for different values of $n_c$ (or equivalently $\phi$) and we store its root $\alpha_m(\phi)$.

This parameter is found to undergo a jump from 0 to >0 at a critical value $\phi_c$ that signals the onset of the nematic transition. To find the binodals we then solve the standard coexistence equations, i.e.

$$\mu_c(\phi_I, f_I) = \mu_c(\phi_N, f_N)$$

$$\Pi_c(\phi_I, f_I) = \Pi_c(\phi_N, f_N)$$

Where $\mu_c = \frac{dF^*}{dN_c} = \frac{d(F_R^*/V)}{dn_c} - \Pi_p \frac{dv}{dn_c}$ is the total chemical potential (evaluated from $F^* = F(\phi, \alpha = \alpha_m)$) and $\Pi_c = -\frac{dF^*}{dV} = \Pi_0 + \Pi_p(v - n_c \frac{dv}{dn_c})$ is the pressure (to compute these we make use of $\frac{N_c}{V} = n_c = \phi/\left(\frac{\pi}{6}D^3 + \frac{\pi}{4}D^2L\right)$). The coexistence equations essentially encode the fact that, in equilibrium, both chemical potential and pressure must be equal in the two (isotropic and nematic) phases.